\documentclass[a4paper,12pt]{article}
\usepackage{epsfig}

\begin{document}

\def\beq{\begin{equation}}
\def\eeq{\end{equation}}
\def\bea{\begin{eqnarray}}
\def\eea{\end{eqnarray}}

\def\dofiga#1#2{\centerline{
\epsfxsize=#1\epsfig{file=#2, width=10cm,height=2cm, angle=0}
\hspace{0cm}
}}
\def\dofigb#1#2{\centerline{
\epsfxsize=#1\epsfig{file=#2, width=8cm,height=10cm, angle=-90}
\hspace{0cm}
}}

\newcommand{\dedouble}{ \stackrel{ \leftrightarrow }{ \partial } }
\newcommand{\deR}{ \stackrel{ \rightarrow }{ \partial } }
\newcommand{\deL}{ \stackrel{ \leftarrow }{ \partial } }
\newcommand{\ci}{{\cal I}}
\newcommand{\ca}{{\cal A}}
\newcommand{\Wp}{W^{\prime}}
\newcommand{\vep}{\varepsilon}
\newcommand{\kk}{{\bf k}}
\newcommand{\pp}{{\bf p}}
\newcommand{\hs}{{\hat s}}
\newcommand{\proj}{\frac{1}{2}\;(\eta_{\mu\alpha}\eta_{\nu\beta}
+  \eta_{\mu\beta}\eta_{\nu\alpha} - \eta_{\mu\nu}\eta_{\alpha\beta})}
\newcommand{\projm}{\frac{1}{2}\;(\eta_{\mu\alpha}\eta_{\nu\beta}
+  \eta_{\mu\beta}\eta_{\nu\alpha}) 
- \frac{1}{3}\;\eta_{\mu\nu}\eta_{\alpha\beta}}

\def\lsim{\raise0.3ex\hbox{$\;<$\kern-0.75em\raise-1.1ex\hbox{$\sim\;$}}} 

\def\gsim{\raise0.3ex\hbox{$\;>$\kern-0.75em\raise-1.1ex\hbox{$\sim\;$}}}

\def\Frac#1#2{\frac{\displaystyle{#1}}{\displaystyle{#2}}}
\def\no{\nonumber\\}
\renewcommand{\thefootnote}{\fnsymbol{footnote}}

\begin{flushright}
CP3-07-37
\end{flushright}

{\Large
\begin{center} 
{\bf 
On the dynamical breaking of chiral symmetry:\\ a new mechanism}
\end{center}
}
\vspace{.3cm}

\begin{center}
Emidio Gabrielli\\
\emph{
Center for Particle Physics and Phenomenology (CP3)\\
Universit\'e catholique de Louvain,
Chemin du Cyclotron, 2\\
B-1348 Louvain-la-Neuve,
Belgium}
\end{center}

\vspace{.3cm}

\hrule \vskip 0.3cm

\begin{center}
\small{\bf Abstract}\\[3mm]
\begin{minipage}[h]{14.0cm}
We consider a U(1) gauge theory, minimally coupled to a massless 
Dirac field, where a higher-derivative term is added to 
the pure gauge sector, as in the Lee-Wick models.
We find that this term can trigger chiral symmetry
breaking at low energy in the weak coupling regime.
Then, the fermion field acquires a mass that turns out to be
a function of both the energy scale associated to the higher-derivative
term and the gauge coupling. The dependence of the fermion mass on the 
gauge coupling is non-perturbative.
Extensions to ${\rm SU}(N)$ gauge theories and fermion-scalar interactions
are also analyzed, as well as to theories with massive gauge fields.
A few implications of these results in the framework of quark-mass 
generation  are discussed.
\end{minipage}
\end{center}

\vskip 0.3cm \hrule \vskip 0.5cm
\vskip 0.3cm
\section{Introduction}
The appearance of indefinite metrics in quantum field theory is often
cause of considerable concern. 
A theory with an indefinite-metric 
contains negative-norm states, namely ``ghosts''.
It is well known that, if ghosts 
belong to the asymptotic states of the S-matrix,
and are also coupled to positive-norm states, unitarity is violated.
However, not all theories with indefinite metric are ill-defined.

As shown long ago by Lee and Wick 
\cite{lw1,lw2}, the introduction of a negative metric in quantum
mechanics does not necessarily spoil unitarity.
It can sometimes lead to a fully unitary S-matrix, provided 
all stable particles in the spectrum have positive square length
(in the Hilbert probability-space).
In other words, if negative norm states have a non-vanishing 
decay width, thus being unstable, they are not among 
the asymptotic states of the S-matrix, and unitarity can be restored.
Problems with the violation of Lorentz invariance,
that might in principle arise due to the indefinite metric
\cite{nakanishi1}, can also be circumvented \cite{lw3}.
A relativistic and unitary S-matrix can indeed be defined
provided the prescription introduced by Cutkosky {\it et al.},
regarding the deformed energy contour in the Feynman integrals, is implemented
\cite{cutk}. Although the above prescription is not
rigorously derived from a Lagrangian field theory approach, 
and might look rather {\it ad hoc} \cite{nakanishi2}, 
it is well defined in perturbation theory \cite{lw3, cutk}.
More recently, non-perturbative formulation of the 
Lee-Wick theories has been analyzed in \cite{lattice}.

The Lee-Wick approach to quantum field theories
prompted the construction of more general theories in which the 
S-matrix is fully unitary, although the Lagrangian is not Hermitian.
Moreover, the exchange of both negative and positive norm states in the
quantum amplitudes turns out to be an advantage.
Ultraviolet divergences may indeed cancel out in the loops 
due to the indefinite metric of the Hilbert space. 

A model satisfying all these requirements
was proposed by Lee and Wick in the framework of quantum electrodynamics 
(QED) \cite{lw2}. 
In particular, by replacing the standard photon field $A_{\mu}$ 
by a complex field $\phi_{\mu}=A_{\mu}+i\, B_{\mu}$ 
in the electromagnetic interaction,
where $B_{\mu}$ is a massive boson field with negative norm, it is possible
to remove all infinities from the electromagnetic mass differences 
between charged particles. This procedure is equivalent to 
the introduction of
a higher (gauge-invariant) derivative term in the Lagrangian of a
primary U(1) gauge field, as can be shown by the 
introduction of auxiliary fields \cite{WLSM}. 
Then, the mass of the ghost field turns out to be 
proportional to the new physics scale $\Lambda$ 
connected to the higher-derivative term.
In order to render charge renormalization finite,
new higher-derivative terms should be introduced for the fermion fields
as well. 

Notice that an analogous mechanism is working in 
the standard Pauli-Villars regularization scheme. In this case, 
the mass of the ghost plays the role of the ultraviolet cut-off.
This ghost decouples from the renormalized theory after its mass
(or, analogously, the Pauli-Villars cut-off) 
is sent to infinity. In other words, the
Lee-Wick approach is to promote the Pauli-Villars cut-off $\Lambda$ 
to a physical mass of the theory.

The ultraviolet behavior of the Lee-Wick theories
is similar to the one of supersymmetric models, as for as the
the cancellation of quadratic divergencies is concerned.
However, while in supersymmetric
models this is achieved by the mutual exchange of virtual particles 
with opposite statistics, in the Lee-Wick theories this is due to
the indefinite metric of the Hilbert space.

Recently, the Lee-Wick model for QED (LWQED) 
has been reconsidered in view of its generalization 
to the Standard Model (SM) of electroweak and strong interactions
\cite{WLSM} (see also \cite{WLSM1} for a few phenomenological applications).
Indeed, this model leads to a theory which is naturally 
free of quadratic divergencies, thus providing an
alternative way to the solution of the hierarchy problem
\cite{WLSM}.
However, contrary to the LWQED model, 
the Lee-Wick SM is not a finite theory, although 
it is still renormalizable.
Higher dimensional operators, containing 
new interactions, naturally appear in higher-derivative theories with 
non-abelian gauge structure.
This leads at most to logarithmic divergencies in the radiative corrections.
Nevertheless, as can be easily understood by power-counting arguments, the 
new higher dimensional operators do not break renormalizability.
This is due to the improved ultraviolet behavior  
of the bosonic propagator $P(k)$ in the deep Euclidean region,
which scales as $P(k) \sim 1/k^4$,
instead of the usual $P(k) \sim 1/k^2$ when $k^2\to \infty$.

Note that, the presence of the energy scale $\Lambda$, associated 
to the higher-derivative term, manifestly breaks (at classical level) 
the conformal symmetry
of the unbroken gauge sector.  Therefore, one may wonder whether this 
term can also 
trigger (dynamically) chiral-symmetry breaking at low energy, or
in other words, whether the fermion field 
could dynamically get a mass $m$ satisfying the condition $m < \Lambda$.
The aim of the present paper is to investigate
this issue by analyzing a general class of
renormalizable models containing higher-derivative terms.

We will show that a non-vanishing mass term for the fermion field can indeed be
generated, depending on the kind of interaction, 
as a solution of the mass-gap equation.
For a massive ghost field, the fermion mass can be predicted, 
and it turns out to be a function of the energy scale $\Lambda$ 
and the gauge coupling constant. 
Moreover, we will see that the dependence of the fermion mass on 
the gauge coupling has a non-perturbative origin.

In section 2, we will consider 
a model where a higher-derivative term is added to 
an exact U(1) gauge theory, 
coupled to a massless fermion field. Then,
we will extend the same approach
to models including renormalizable scalar(pseudoscalar)-fermion interactions
with massless scalar(pseudoscalar) fields.
In section 3, we extend the analysis of section 2
to the case of ${\rm SU}(N)$ gauge interactions
in the presence of a higher-derivative term for the non-abelian gauge fields.
In section 4, the same mechanism is analyzed for the
case of interactions mediated by a massive gauge field. 
Our conclusions are given in section 5.

\section{Abelian gauge and scalar interactions}
We start our analysis by considering a minimal version of the
Lee-Wick extension of the U(1) gauge theory \cite{lw2}, where
a (gauge-invariant) dimension-6 operator containing higher-derivatives
is added to the free Lagrangian of the U(1) gauge
sector. The corresponding gauge field $A_{\mu}$ is then minimally coupled to a 
Dirac field $\psi$.

In contrast to the original Lee-Wick model \cite{lw2},
we do not impose here finite charge renormalization, since this would
require the introduction of extra higher-derivative terms 
in the fermion sector.
Since we are interested in the dynamical fermion-mass generation,
we will switch off the bare-mass of the fermion field.
We consider the following gauge-invariant Lagrangian ${\cal L}$
\bea
{\cal L}=-\frac{1}{4}\, F_{\mu\nu}\, F^{\mu\nu}
\,+\,\frac{1}{\Lambda^2} 
\left(\partial^{\alpha} F_{\alpha\mu}\right)
\left(\partial^{\beta} F_{\beta}^{\mu}\right)\,+\, 
i\bar{\psi}\gamma_{\mu}D^{\mu}\psi\, ,
\label{Lag}
\eea
where the field strength and the covariant derivative are
defined as $F_{\mu\nu}=\partial_{\mu}A_{\nu}-\partial_{\nu}A_{\mu}$ and 
 $D_{\mu}=\partial_{\mu}+i g A_{\mu} $ respectively.

In order to make the theory consistent in perturbation theory, 
it is necessary to add a gauge-fixing term. We choose
the covariant term 
$L_{\xi}=-(\partial_{\mu}A^{\mu})^2/(2\xi)$, where $\xi$, as usual, plays the 
role of the gauge-fixing parameter. 
Due to the abelian structure of the theory, no Faddeev-Popov 
ghosts are required in this case.

According to the above gauge-fixing term, 
the propagator of the gauge field $A_{\mu}$ in momentum space is given by
\bea
D_{\mu\nu}(k)=\frac{-i\Lambda^2}{k^2\left(\Lambda^2-k^2\right)}
\left(\eta_{\mu\nu}-(1-\xi)\frac{k_{\mu}k_{\nu}}{k^2}-
\xi\, \frac{k_{\mu}k_{\nu}}{\Lambda^2}\right)\, .
\label{prop}
\eea
This propagator has two poles: at $k^2=0$ and $k^2=\Lambda^2$. 
Indeed, the Lagrangian in Eq.(\ref{Lag}) describes
two independent (on-shell) spin-1 fields:
massless one and massive one, with positive and negative norm
respectively. As shown in \cite{WLSM},
this can be easily shown by the help of an (on-shell) auxiliary field which
can linearize the Lagrangian in Eq.(\ref{Lag}).
However, here it is more convenient to 
work in the representation of the gauge field $A_{\mu}$
as given in Eq.(\ref{Lag}), with the propagator as in Eq.(\ref{prop}).
The Feynman rules for the coupling of the Dirac field $\psi$ to the 
gauge field $A_{\mu}$ are the same as in the standard U(1) gauge theory.

Since no mass term for the fermion field is present at tree-level,
the Lagrangian in Eq.(\ref{Lag}) is also
invariant under global chiral transformations, namely
$\psi\to e^{i\gamma_5  \varepsilon}\psi$, where $\varepsilon$ 
is a constant parameter.
However, due to the presence of the scale $\Lambda$, 
the conformal symmetry is broken at classical level.
Then, one might wonder whether $\Lambda$
could also trigger chiral-symmetry breaking at low energy, 
hence dynamically generating a mass term for the fermion field
lower than $\Lambda$.

In order to answer this question, 
we consider below the equation for the fermion 
mass-gap following the approach of the
Nambu-Jona-Lasinio (NJL) model \cite{NJL}. 
One has then to calculate 
the one-loop contribution to the fermion self-energy as a function of
the fermion pole mass $m$. Due to the improved asymptotic 
behavior of the gauge propagator ( $\sim 1/k^4$ when $k^2\to \infty$,
as induced by the presence of the ghost field), this contribution
turns out to be finite. Therefore, the fermion mass can be predicted 
as a function of the gauge coupling and $\Lambda$. 
Analogous results can be obtained by implementing 
the Pauli-Villars regularization, although their
physical interpretation is different in that case.

\begin{figure}[tpb]
\dofiga{3.1in}{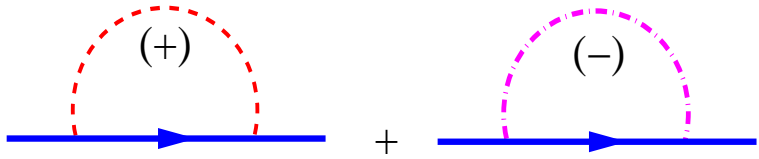}
\vspace{-.3cm}
\caption
{\small Feynman diagrams contribution to the fermion self-energy
at one-loop in the U(1) gauge theory with a higher-derivative term.
The symbols $(+)$ and $(-)$ 
indicate the usual gauge propagator (dashed line) 
and the  
``massive-ghost'' propagator (dashed-dot line) respectively, while
the continuous line stands for the fermion propagator.
}
\label{Fig1}
\end{figure}

The Feynman 
diagrams contribution to the one-loop self-energy are shown in Fig. 
\ref{Fig1}.
By assuming the propagator in Eq.(\ref{prop}), we obtain
\bea
\Sigma(\hat{p},m)= \frac{\alpha}{2\pi}\,
\left(\int_0^1\, dx \left(2m-x\, \hat{p}\right)\, \log\left(
\frac{x\Lambda^2+(m^2-p^2x)(1-x)}{(m^2-p^2x)(1-x)}\right)\, + I(\xi)
\right)
\label{selfE}
\eea
where $\alpha= g^2/(4\pi)$ and $\hat{p}\equiv \gamma^{\mu} p_{\mu}$. 
The integral $I(\xi)$ in Eq.(\ref{selfE}) contains
the pure gauge-dependent contribution to the self-energy 
due to terms proportional to $k_{\mu}k_{\nu}$
in the gauge propagator. However,
$I(\xi)$ vanishes on shell ($\hat{p}=m$), since 
the self-energy when evaluated on the physical pole 
is a gauge invariant quantity. As we will see later on, this property 
is not spoiled by an explicit mass term for the gauge
field, since the (on-shell) external current is always conserved in 
an abelian gauge theory.
Being interested in the self-energy evaluated on-shell, the explicit 
expression of $I(\xi)$ is not needed here.

The exact resummation at any order of the 
self-energy contribution to the fermion propagator $S(p)$ gives
the well-known result 
\bea
S(p)=\frac{i}{\hat{p}-m_0-\Sigma(\hat{p},m)}\, .
\label{S}
\eea
In the above formula $m_0$ is the tree-level bare mass, 
while, as explained above, $m$ is the physical pole mass.
Since we are interested in the dynamical generation
of fermion mass, following the approach of NJL \cite{NJL}, we
set in Eq.(\ref{S}) the bare mass $m_0$ to zero.
Then, one obtains the well known
self-consistent equation for the mass-gap \cite{NJL}
\bea
m=\Sigma(\hat{p},m)|_{\hat{p}=m}\, .
\label{gap-eq}
\eea
Notice that, both  Eq.(\ref{gap-eq}) and its solutions are
gauge invariant.

By substituting the on-shell condition $\hat{p}=m$ in 
Eq.(\ref{selfE}) we obtain 
\bea
m&=& -\frac{\alpha}{2\pi}\,m\, \int_0^1\, dx\, (2-x)\, 
\log{\left(\frac{m^2}{\Lambda^2}\frac{(1-x)^2}{x}\right)}\, +{\cal O}(m^2/\Lambda^2)
\label{selfE1_gen}
\eea
Terms of the order ${\cal O}(m^2/\Lambda^2)$ are neglected, since
we are interested in finding non-trivial solutions of Eq.(\ref{selfE}) 
corresponding to the breaking of chiral symmetry at low energy, that is
at $m\ll \Lambda$.
Notice that, due to the chiral symmetry of the original Lagrangian,
there is always a trivial solution of Eq.(\ref{selfE1_gen}) 
corresponding to the case $m=0$. Remarkably, as in the NLJ model \cite{NJL},
Eq.(\ref{selfE1_gen}) 
allows also a non-trivial solution with $m\neq 0$, satisfying
the condition $m < \Lambda$.
According to the interpretation of the NJL equation, 
the non-trivial solution should be identified with the physical one, 
since it corresponds to the non-perturbative (true) vacuum of the theory.
On the other hand, the massless solution is always present, being connected
to the false vacuum of perturbation theory, where chiral symmetry is unbroken.

Then, solving the self-consistent equation for the mass-gap
\bea
1&=& -\frac{\alpha}{2\pi}\, \int_0^1\, dx\, (2-x)\, 
\log{\left(\frac{m^2}{\Lambda^2}\frac{(1-x)^2}{x}\right)}\, +
{\cal O}(m^2/\Lambda^2)\, ,
\label{selfE1}
\eea
we obtain
\bea
m=\Lambda \exp{\left[-\frac{2\, \pi}{3\, \alpha}+\frac{1}{4}\right]}\, .
\label{sol1}
\eea
This solution corresponds to a dynamical breaking of the chiral symmetry.
The $1/\alpha$ dependence in Eq.(\ref{sol1}) shows
the non-perturbative origin of such effect.
The fact that there exists a non-trivial solution to the mass-gap equation
is a peculiar property of unbroken gauge theories. 
Indeed, as we will see in the following, there is no solution 
satisfying the condition $m < \Lambda$ in the 
case of fermion-scalar(pseudoscalar) interactions.

On the other hand, Eq.(\ref{sol1}) is not yet the complete solution,
since the resummation of the 
leading-Log terms $(\alpha\log{(\Lambda/m}))^n$, arising from 
the inclusion of vacuum polarization diagrams, should be 
taken into account.
From the renormalization group equation, we known that all these effects 
can be re-absorbed in the running coupling constant $\alpha(Q)$ evaluated at
the scale $Q\sim m$. This effect then can be taken into account 
by replacing $\alpha\to \alpha(m)$ in Eq.(\ref{sol1})
\bea
m=\Lambda \exp{\left[-\frac{2\, \pi}{3\, \alpha(m)}+\frac{1}{4}\right]}\, ,
\label{solnew}
\eea
where $\alpha(m)$ is related to $\alpha(\Lambda)$  by
\bea
\alpha(m)=\frac{\alpha(\Lambda)}{1+b\alpha(\Lambda)\log{(\Lambda/m)}}
\label{alpha_run}
\eea
with $b=2/(3\pi)$ for a U(1) gauge theory. 

Let's now consider the case of $N_f$ fermions having
the same charge $e$. Then,
Eq.(\ref{solnew}) can be easily generalized to a set of $N_f$
self-consistent equations, as follows
\bea
m_f=\Lambda \exp{\left[-\frac{2\, \pi}{3\, \alpha(m_f)}+\frac{1}{4}\right]}\, .
\label{solnew1}
\eea
Although Eq.(\ref{solnew1}) 
depends on  $m_f$ through the running coupling 
constant $\alpha(m_f)$, its solution is consistent with 
a degenerate spectrum only.
This is just a consequence of charge universality. 
In order to prove that, let's assume 
a non-degenerate spectrum for the solution of Eq.(\ref{solnew1}).
Then, one can order the mass convention according to $m_f> m_{f-1}$, where
$f=1,\dots, N_f$. Now we rewrite the 
$\alpha(m_{f})$ dependence in terms of  $\alpha(m_{f-1})$
into Eq.(\ref{solnew1}), and we get 
$\left(\frac{m_f}{m_{f-1}}\right)^{1-C}=1$,
where $C=4/9 N_{f-1}$, which implies $m_f=m_{f-1}$. Iterating
the same procedure for $m_{f-1}$ and $m_{f-2}$, we
find $m_{f}=m_{f-1}=m_{f-2}$, and hence a degenerate spectrum.

Since Eq.(\ref{solnew1}) has been derived within perturbation theory,
we have to check that its solution is consistent with the perturbative regime.
In particular, one can require that the coupling constant $\alpha$
is perturbative up to the $\Lambda$ scale and positive, that is
$0< \alpha(\Lambda) < 1$.
To this end, we have to first express $m_f$ in Eq.(\ref{solnew1}) 
in terms of $\alpha(\Lambda)$, obtaining
\bea
m_f=\Lambda\exp\left[\left(-\frac{2\pi}{3\alpha(\Lambda)}+\frac{1}{4}\right)
\left(\frac{9}{9-4N_f}\right)\right]
\label{sol_univ}
\eea
for any $f$. Remarkably, 
Eq.(\ref{sol_univ}) is compatible with the condition 
$m_f < \Lambda$ and the weak coupling regime ( $\alpha(\Lambda) \ll 1 $),
provided the number of charged fermions is $N_f\le 2$. If $N_f >2$, then
the condition $m_f < \Lambda$ is satisfied only for $\alpha(\Lambda)<0$,
which is clearly inconsistent. Nevertheless, we will see in the next section 
that the constraint $N_f\le 2$ can be relaxed for
non-abelian gauge interactions.

Now we consider the most general case of $N_f$ fermions 
$f$ minimally coupled to 
a U(1) gauge theory as in Eq.(\ref{Lag}), 
where all charges are different, namely $Q_f$ in unity of $e$. 
Then, the non-universality of the charges
can remove the degeneracy of the spectrum.
Eq.(\ref{sol_univ}) can be easily generalized to this case
of non-universal charges. The corresponding set 
of $N_f$ self-consistent equations is given by
\bea
m_f=\Lambda\exp\left[\left(-\frac{2\pi}{3\alpha(\Lambda)}+
\frac{Q^2_f}{4}\right)
\left(\frac{9}{9Q^2_f-4\sum_f Q_{f}^2}\right)\right]\, ,
\label{sol_nouniv}
\eea 
where $\alpha=e^2/(4\pi)$. 
We can see that, if
$Q_f > Q_{f-1}$, then $m_f> m_{f-1}$, provided 
$\alpha(\Lambda) < 8\pi/(3Q_{\rm max}^2)$, where $Q_{\rm max}$ is the
largest charge.
A closer inspection of Eq.(\ref{sol_nouniv}) shows that,
also in this case, the constraint $N_f \le 2$ cannot be avoided. 
Indeed, let's consider the contribution to the 
lowest mass eigenvalue. In case we order the charges
as $|Q_i| < |Q_j|$ for $i< j$ (where $(i,j)=1,\dots,N_f$), this
would correspond to the fermion with charge $Q_1$.
Moreover, the requirement of positivity of the second term in 
parenthesis in the exponential would imply that
$Q^2_1> 4/5\sum_{i=2}^{N_f} Q_{i}^2$. This last condition cannot be 
satisfied if $N_f > 2$, since $|Q_1| < |Q_i|$ for $i \ge 2$. Therefore, also 
in the general case of non-universal charges the restriction $N_f \le 2$ holds,
and is supplemented by the additional constraint $1<Q^2_2/Q_1^2<5/4$.

Let's now consider the case of a chiral model with massless
scalar(pseudoscalar) fields coupled to a massless fermion field.
In order to implement chiral symmetry, 
the minimal number of real scalar fields required is two,
namely a scalar ($\varphi$) and pseudoscalar ($\bar{\varphi}$) field.
In analogy with the U(1) gauge theory discussed above, we add
a higher derivative term in the scalar sector.
The corresponding Lagrangian is
\bea
{\cal L}=L_0(\varphi)+L_0(\bar{\varphi})+L_0(\psi)-
\frac{1}{2\Lambda^2}\left((\partial_{\mu}\partial^{\mu}\varphi)^2
+(\partial_{\mu}\partial^{\mu}\bar{\varphi})^2\right)
+\lambda\left(\bar{\psi}\psi\varphi+i\bar{\psi}\gamma_5\psi\bar{\varphi}
\right)
\label{Lscalar}
\eea
where $L_0(i)$ stands for the canonical kinetic term
of the corresponding fields $i=\varphi,\bar{\varphi},\psi$.
The r.h.s of Eq.(\ref{Lscalar})
is invariant under global chiral transformations, 
that, for an infinitesimal parameter $\varepsilon$, are defined as
\bea
\delta \psi=-i\varepsilon \gamma_5 \psi,~~~
\delta \bar{\psi}=-i\varepsilon \bar{\psi}\gamma_5,~~~
\delta \varphi=-2\varepsilon \bar{\phi},~~~ 
\delta \bar{\varphi}=2\varepsilon \phi\, .
\eea
The Feynman diagrams for the fermion self-energy are the same as in 
Fig. \ref{Fig1}, where the gauge propagator is replaced by the 
corresponding scalar and pseudoscalar fields propagator.
Then, one gets the following result for the self-consistent equation 
\bea
1&=& \frac{\alpha_{\lambda}}{2\pi}\, \int_0^1\, dx\, x\, 
\log{\left(\frac{m^2}{\Lambda^2}\frac{(1-x)^2}{x}\right)}
+{\cal O}(m^2/\Lambda^2)\, ,
\label{selfE2}
\eea
where
$\alpha_{\lambda}\equiv \lambda^2/(4\pi)$. The resummation of 
the leading-Log terms 
$(\alpha_{\lambda} \log{(\Lambda/m)})^n$, as discussed above,
can again be reabsorbed
in the running coupling constant evaluated at the scale $m$. In particular,
the replacement $\alpha_{\lambda}\to \alpha_{\lambda}(m)$ 
should be done in Eq.(\ref{selfE2}).
In this case, the sign of the term proportional to 
$\log{(m/\Lambda)}$ in Eq.(\ref{selfE2}) is opposite
with respect to Eq.(\ref{selfE1}).
This has dramatic consequences for chiral symmetry breaking.
The non-trivial solution of Eq.(\ref{selfE2}) would imply 
\bea
m=\Lambda \exp{\left[\frac{2\, \pi}{\alpha_\lambda(m)}+
\frac{5}{4}\right]}\, ,
\label{sol2}
\eea
which is inconsistent with the requirement $m < \Lambda$.
We conclude that, contrary to the gauge interactions case, 
chiral symmetry breaking cannot be triggered 
at low energy (that is for $m < \Lambda$) by means of 
pure fermion-scalar(pseudoscalar) interactions as in Eq.(\ref{Lscalar}).

We stress here that all these results 
hold under the assumption that the Lee-Wick
theories can be well-defined at non-perturbative level.
Although there is not yet any rigorous proof that the Lee-Wick extension
could also work at non-perturbative level, there are studies
in this direction leading to a consistent non-perturbative 
approach on the lattice \cite{lattice}.

\section{${\bf SU}(N)$ gauge interactions}
In this section 
we generalize the U(1) gauge model, presented in section 2, to the
non-abelian SU$(N)$ gauge theory. We add to the standard
Yang-Mills Lagrangian the corresponding
(gauge-invariant) higher-derivative term for the non-abelian gauge fields.
Following the Lee-Wick generalization of the pure ${\rm SU}(N)$ sector 
\cite{WLSM}, we consider the following Lagrangian
\bea
{\cal L}\, =\, -\frac{1}{2}{\rm Tr}\left[ \hat{F}_{\mu\nu}
\hat{F}^{\mu\nu}\right]\,+\,\frac{1}{\Lambda^2}{\rm Tr}\left[ 
\left(\hat{D}^{\alpha}\hat{F}_{\alpha\mu}\right)
\left(\hat{D}^{\beta}\hat{F}_{\beta}^{\, \mu}\right)\right]\,+\,
i\bar{\psi}\gamma_{\mu}\hat{D}^{\mu}\psi \, ,
\label{lagSUN}
\eea
where $\hat{F}^{\mu\nu}$ is the standard field strength of Yang-Mills theories,
and $\hat{D}_{\mu}=\partial_{\mu}+ig \sum_a T^a A^a_{\mu}$, with $T^a$
the ${\rm SU}(N)$ generators in the fundamental representation.
The trace ${\rm Tr}$ is understood acting on the fundamental representation of
the $T^a$ matrices.
For the linear decomposition of the Lagrangian in Eq.(\ref{lagSUN}) in terms
of an auxiliary field, see \cite{WLSM}.

If we add to the Lagrangian in Eq.(\ref{lagSUN})
the covariant gauge-fixing term 
${\cal L}_{GF}=-{\rm Tr}[(\partial_{\mu}\hat{A}^{\mu})^2/2\xi]$, 
the gauge propagator is given by 
\bea
D_{\mu\nu}^{ab}(k)=\delta^{ab}\frac{-i\Lambda^2}{k^2\left(\Lambda^2-k^2\right)}
\left(\eta_{\mu\nu}-(1-\xi)\frac{k_{\mu}k_{\nu}}{k^2}-
\xi\, \frac{k_{\mu}k_{\nu}}{\Lambda^2}\right)\, ,
\label{prop_SUN}
\eea
where the indices $a,b$ run on the adjoint representation of ${\rm SU}(N)$.
Notice that, apart from the $\delta^{ab}$ term, 
this is the same propagator as in Eq.(\ref{prop}).

The calculation of the self-consistent equation for the mass-gap 
proceeds as in the case of the U(1) gauge theory. 
At one-loop, the contribution to the 
fermion self-energy is provided by the same kind of diagrams as in 
Fig. \ref{Fig1}.
As already mentioned, since the fermion self-energy 
evaluated on-shell is gauge invariant, the contribution of the gauge 
propagator proportional to terms $k_{\mu}k_{\nu}$ vanishes. 
Therefore, the expression of the self-energy at one-loop 
is the same as in Eq.(\ref{selfE}), 
apart from the ${\rm SU}(N)$ factor $C_F=(N^2-1)/(2N)$. 
Then, the corresponding solution to the self-consistent equation is
\bea
m=
\Lambda\exp\left[-\frac{2\pi}{3\alpha(m) C_F}+\frac{1}{4}\right]\, .
\label{sol1_SUN}
\eea
As in the U(1) case, $m$ has now to be expressed
in terms of the running coupling coupling $\alpha(\Lambda)$ evaluated at
the high-scale $\Lambda$.
Apart from the factor $C_F$, there is now a crucial difference 
in the effective coupling constant.
The opposite sign in the $\beta$-function for a SU($N$) gauge theory,
with respect to the abelian case, has dramatic impact
on the solution. In particular, the constraint 
on the total number of charged fermions will be strongly relaxed.
We stress that, since the scale $\Lambda$ is assumed to be much
larger than the characteristic intrinsic scale associated to ${\rm SU}(N)$, 
that we name $\Lambda_{{\rm SU}(N)}$ ($\Lambda_{{\rm SU}(3)}=\Lambda_{QCD}$
in the QCD case)\footnote{
Anyhow, one should keep in mind that for a generic ${\rm SU}(N)$ 
gauge theory the fundamental scale $\Lambda_{{\rm SU}(N)}$ 
is a free parameter.}, the massive ghost-field does not contribute to the 
${\rm SU}(N)$ $\beta$-function below the $\Lambda$ scale.
However, above the $\Lambda$ scale, the one-loop $\beta$-function is modified 
due to the contribution of the massive ghost field. This effect has been 
recently evaluated in \cite{betaSUN}.

If we substitute in Eq.(\ref{sol1_SUN}) the running coupling $\alpha(m)$
in Eq.(\ref{alpha_run}) with the corresponding coefficient 
$b=-1/(6\pi)\left(11N-2N_f\right)$ of the SU($N$) $\beta-$function, we get
\bea
m=\Lambda\exp\left[\left(-\frac{6\pi}{\alpha(\Lambda)}
+\frac{9C_F}{4}\right)
\frac{1}{9C_F+11N-2N_f}
\right]\, .
\label{solSUN}
\eea
Now, one can rearrange Eq.(\ref{solSUN}) in a more compact expression.
If we substitute 
$\alpha(\Lambda)=6\pi/[(11N-2 N_f)\log(\Lambda/\Lambda_{{\rm SU}(N)})]$ 
in Eq.(\ref{solSUN}), where $\Lambda > \Lambda_{{\rm SU}(N)}$, we obtain
\bea
m=\Lambda\left(\frac{\Lambda_{{\rm SU}(N)}}
{\Lambda}\right)^{\beta}\, e^{\gamma}\, ,
\label{mSUN}
\eea 
where the coefficients $\beta$ and $\gamma$ are given by
\bea
\beta=\frac{11N-2N_f}{9C_F+11N-2N_f}\, ,~~~~~~
\gamma=\frac{9C_F}{4\left(9C_F+11N-2N_f\right)}\, .
\eea
In the QCD case with $N_f=6$, we have $\beta=7/11$ and $\gamma=1/11$.

\begin{figure}[tpb]
\dofigb{3.1in}{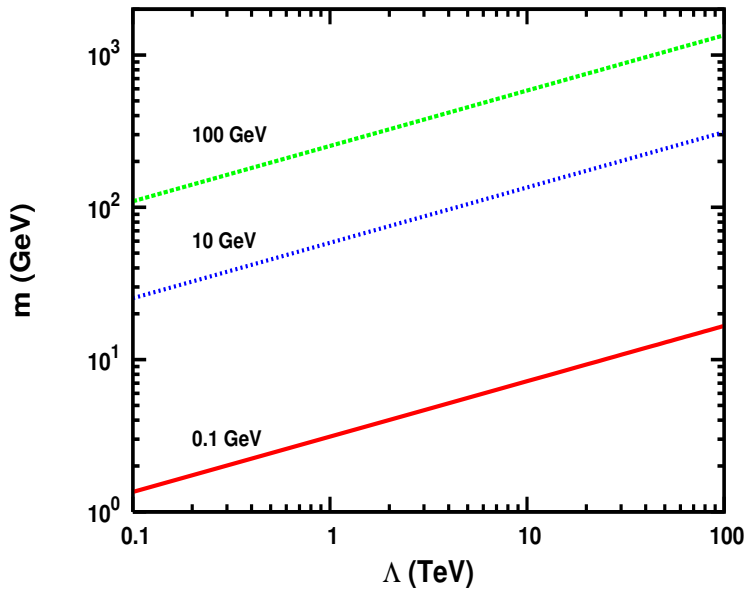}
\vspace{0.1cm}
\caption
{\small The fermion mass $m$ (in GeV) versus the higher-derivative scale 
$\Lambda$ (in TeV) in the SU$(3)$ case, with fermions number
$N_f=6$, and for the values $\Lambda_{{\rm SU}(3)}=0.1,\, 10,\, 100$ GeV.}
\label{Fig2}
\end{figure}

In Fig. \ref{Fig2} we show the dependence of $m$ versus $\Lambda$ in
the SU(3) case with $N_f=6$, for a few $\Lambda_{{\rm SU}(3)}$ values,
$\Lambda_{{\rm SU}(3)}=0.1,\, 10,\, 100$ GeV, where the first one
should roughly correspond to the QCD case.
We can see that, a higher-dimensional term
with a scale of the order of $\Lambda=1$ TeV
would generate in QCD a fermion mass of the order of $m \sim 3$ GeV.

We consider now some potential phenomenological
application of this mechanism. The fact that scalar-fermion
interactions act with a term of
opposite sign in the argument of the exponential
in Eq.(\ref{sol2}) with respect to Eq.(\ref{solSUN}),
suggests that the presence of both Yukawa and
gauge couplings could induce a fermion mass-splitting.
Remarkably, due to the non-perturbative dependence of the fermion 
mass on the Yukawa coupling (cf. Eq.(\ref{sol2})), 
a large mass-splitting could be achieved 
with Yukawa interactions of similar strength.

On the other hand, 
this mechanism requires scalar and pseudoscalar fields with positive
norm that are massless.
In order to have a realistic viable (and alternative to the standard 
Higgs mechanism) mechanism for quark mass generation, 
one needs {\it confined} massless scalar and pseudoscalar fields.
Then, due to their Yukawa couplings with quarks, scalars and pseudoscalars
must be in the adjoint representation of color, which is a quite intriguing 
phenomenological possibility. 

Although attractive, we remind that, at this stage, this 
is just a speculative idea.
A realistic model of fermion mass generation in the SM would 
require a careful analysis, which is beyond the purpose of the present paper.

\section{Massive vector field}
In this section, we study whether the proposed
mechanism for chiral symmetry breaking can work 
for short-distance interactions,
by considering a massive vector field coupled to a 
conserved fermion current. 
We start from the same model as in Eq.(\ref{Lag}), with the 
addition of a mass term for the gauge field. The corresponding Lagrangian
is then
${\cal L}^{\prime}={\cal L} + \frac{1}{2}M^2 A_{\mu} A^{\mu}$, where 
${\cal L}$ is given in Eq.(\ref{Lag}).
We restrict our analysis to the case of a single fermion field. 
The generalization
to an arbitrary number of fermions will be straightforward.

No matter which mechanism generates the gauge
field mass, the gauge propagator, in the unitary gauge, is given by
\bea
D_{\mu\nu}(k)=\frac{-i\Lambda^2}{\left(k^2-M^2\right)
\left(\Lambda^2-k^2\right)}
\left(\eta_{\mu\nu}-\frac{k_{\mu}k_{\nu}}{M^2}\right)\, ,
\label{propm}
\eea
where we assume that $\Lambda$ is the highest scale in the theory,
namely $\Lambda \gg M$. Notice that, $\Lambda > 2M$
is necessary in order to have a non-vanishing
ghost-field decay width. Indeed, in case fermions get
a mass $m > M$, the ghost could anyway decay in two gauge bosons
(i.e., by means of one-loop fermion diagram).

Then, Eq.(\ref{selfE1}) for the mass-gap becomes
\bea
1= -\frac{\alpha}{2\pi}\, \int_0^1\, dx\, (2-x)\, 
\log{\left(\frac{x M^2+m^2 (1-x)^2}{x \Lambda^2+m^2(1-x)^2}\right)}\, .
\label{selfEM}
\eea
The r.h.s of Eq.(\ref{selfEM}) is insensitive, 
at one-loop, to the 
dynamics which generates the gauge boson mass. Indeed, as 
explained in section 2, due to the conservation 
of the  on-shell fermion current, the $k_{\mu}k_{\nu}$ 
contribution in the gauge propagator to the on-shell
self-energy vanishes.

The new mass scale $M$ in the one-loop self-energy makes
the corresponding solution of 
Eq.(\ref{selfEM}) dependent on the ratio $M/\Lambda$.
The exact analytical expression is quite difficult to obtain,
due to the non-trivial dependence of  Eq.(\ref{selfEM})
on $m$, $M$, and $\Lambda$.
Nevertheless, there are three particular scenarios where the problem can be 
easily solved by making use of some perturbative expansion:
i) $M \ll m\ll \Lambda$, ii) $m\simeq M\ll \Lambda$, iii) 
$m\ll M\ll \Lambda$.

In the first case ($m\gg M$), one can see that there exists always a 
non-trivial solution for $m\neq 0$.
Indeed, the leading contribution to $m$
would be the same as in Eq.(\ref{sol1}), 
apart from small corrections of the  order of ${\cal O}(M/\Lambda)$. 
The second and third cases admit different non-trivial solutions
with respect to Eq.(\ref{sol1}) and we are going to analyze them in the
following.

Let's starts with the case $m\simeq M$. Assuming $m=M$ in 
the integral in Eq.(\ref{selfEM}), and solving the equation for $m$, we get
\bea
m=M=\Lambda\, \exp\left[-\frac{2\pi}{3\alpha}+\frac{5-2\sqrt{3}\pi}{12}\right]
\, ,
\label{sol_meqM}
\eea
where the last term in the exponential can be well approximated by
$-1/2$ (indeed, $\frac{5-2\sqrt{3}\pi}{12}\simeq -0.49$).
As previously shown for U(1) and SU($N$) gauge interactions,
in order to resum the leading-Log terms, $\alpha$ should be replaced
by a  running coupling constant evaluated at the characteristic 
renormalization scale of the problem. 
In the following we will omit this dependence in the notation.

Let's now consider the iii) scenario, where the 
fermion is the lightest state, namely 
$m\ll M\ll \Lambda$.
By expanding the integral in Eq.(\ref{selfEM}) in terms of
$m/\Lambda \ll 1$, $M/\Lambda \ll 1 $, 
and  $m/M \ll 1$, and retaining only the leading-Log contributions, 
we find that there exists a non-trivial solution for $m$ 
provided the following equation holds
\bea
\frac{2\pi}{\alpha}+3\log{\left(\frac{M}{\Lambda}\right)}
-\frac{4m^2}{M^2}\log{\left(\frac{m}{\Lambda}\right)}=0\, .
\label{sol3}
\eea
We see  that Eq.(\ref{sol3}) can be satisfied only if the term
$3\log{\left(\frac{M}{\Lambda}\right)}$ can compensate the
large term $\frac{2\pi}{\alpha}$. This, for $m\neq 0$,
requires a critical $M$ value.
An approximate analytical solution for $m$ can be obtained by using 
the following ansatz for $M$
\bea
M=\Lambda\exp\left[-\frac{2\pi}{3\alpha}-\frac{\delta^2}{3}\right]
\label{ansatz}
\eea
where $\delta$ is a (real) free dimensionless 
parameter satisfying the condition 
$0< \delta \ll \sqrt{3/2}$. Notice that, in order to
satisfy Eq.(\ref{sol3}) one must assume a considerable tuning between
$M$ and $\Lambda$ as shown by the dependence of $\delta$ in the
argument of the exponential. By substituting 
Eq.(\ref{ansatz}) in Eq.(\ref{sol3}), we find 
\bea
m\simeq \sqrt{\frac{3\alpha}{8\pi}}\,\delta\,M\, .
\label{mass_sol}
\eea
Eq.(\ref{mass_sol})  is quite different from the U(1) massless case
in Eq.(\ref{sol1}).
The fermion mass here is suppressed by a term 
proportional to $(\sqrt{\alpha}\delta) \ll 1$.
Indeed, as expected from Eq.(\ref{sol3}), 
when $M \gg \Lambda \exp\left[-\frac{2\pi}{3\alpha}\right]$, no 
$m\neq 0$ satisfying the condition $m \ll M$ can be obtained.

Summarizing, we get the following results
for $m$, as a function of $M/\Lambda$, in the three difference
ranges
\bea
&{\bf i)}&~~0\,<\, \frac{M}{\Lambda} \, \ll 
\exp\left[-\frac{2\pi}{3\alpha}-\frac{1}{2}\right]
~~\Rightarrow~~
m\simeq \Lambda \exp\left[-\frac{2\pi}{3\alpha} +\frac{1}{4}\right]\, ,
\nonumber \\
&{\bf ii)} &~~\exp\left[-\frac{2\pi}{3\alpha}-\frac{\delta^2}{3}\right]\, \lsim\, 
\frac{M}{\Lambda}\, \lsim\, 
\exp\left[-\frac{2\pi}{3\alpha}\right]
~~\Rightarrow~~
m\simeq \sqrt{\frac{3\alpha}{8\pi}}\,\delta M\, ,
\nonumber \\
&{\bf iii)} &~~\exp\left[-\frac{2\pi}{3\alpha}\right] \, \ll \, 
\frac{M}{\Lambda} \,<\, 1~~\Rightarrow~~
m=0\, ,
\eea
where $0< \delta \ll \sqrt{3/2}$. These are approximate results.
A precise $m$ value at the thresholds would require the knowledge 
of the exact analytical expression of $m$ in terms of $M/\Lambda$.

In conclusion, in order to trigger
chiral symmetry breaking at low energy in the weak coupling
regime, the gauge boson mass should
satisfy a critical condition. In particular, in the case of U(1) gauge
interactions, one needs 
$M < \Lambda \exp\left[-\frac{2\pi}{3\alpha}\right]$,
where $\alpha$ is understood to be evaluated at the scale $M$.
However, a strong tuning between $M$ and $\Lambda$ is required
in order to obtain a fermion mass $m\ll  M$.

\section{Conclusions}
We have analyzed a new mechanism for chiral-symmetry-breaking
which is based on renormalizable models with higher-derivatives.
As an example, we considered chiral theories
with gauge-fermion and scalar-fermion interactions,
where a higher-derivative term is added to the free bosonic sector 
of the Lagrangian, as in the Lee-Wick models.

As order-parameter of chiral symmetry breaking,
we considered the fermion mass-gap $m$.
The corresponding self-consistent equation for the mass-gap has been 
derived by using the approach of the NJL model \cite{NJL}.
Remarkably, we find that a non-trivial solution $m\neq 0$
exists in the weak coupling regime, satisfying the condition $m\ll \Lambda$.
We show that this is
a peculiar property of the fermion-gauge interactions, and does not
hold in the case of (pure) scalar-fermion interactions.
Moreover, due to the presence of the
ghost field, the contribution to the fermion self-energy is finite,
and the mass-gap can be predicted. Then, $m$ turns out to be a 
function of the higher-derivative scale $\Lambda$ and the gauge coupling
constant. Although the self-consistent equation has been derived 
in a perturbative regime, the mass dependence on 
the gauge coupling is a pure non-perturbative effect.

We generalized this mechanism to the SU($N$) gauge interactions by adding
the corresponding higher-derivative term to the non-abelian gauge fields. 
We find that there exists a non-vanishing 
fermion mass-gap in the weak coupling regime,
provided $\Lambda> \Lambda_{{\rm SU}(N)}$.
In particular, the mass-gap turns out to be a simple function of
 $\Lambda$ and $\Lambda_{{\rm SU}(N)}$, namely
$m=\Lambda\left(\frac{\Lambda_{{\rm SU}(N)}}
{\Lambda}\right)^{\beta}\, e^{\gamma}$, where $\beta$ and $\gamma$ are
some coefficients depending on $N$ and fermions number $N_f$.
In the SU(3) case, with $N_f=6$, we get for these coefficients
$\beta=7/11$ and  $\gamma=1/11$. We think that 
potential lattice studies of these theories could be very 
helpful in testing the above results on the fermion mass-gap.

We also considered the same mechanism in the presence of a massive gauge
field. Then, we show that in order to trigger chiral symmetry breaking
at low energy, the mass $M$ of the gauge field needs to satisfy 
a critical condition. However,
a strong tuning between $M$ and $\Lambda$ is required
in order to obtain a fermion mass $m\ll  M$.

In conclusion, we believe that further studies are necessary
to assess the real potential of this mechanism.
In particular, it would be interesting to analyze
the interplay of SU($N$) gauge-fermion and scalar-fermion interactions with 
higher-derivative terms.
As discussed in section 3, this might help in explaining 
the observed quark spectrum in a natural way. This would require
scalar fields in the adjoint representation of color, which is a quite
intriguing phenomenological possibility. 

We think that this mechanism for the fermion mass generation
would deserve some consideration also in the framework of technicolor models.

\newpage
\section{Acknowledgments}
We thank Fabio Maltoni for useful discussions 
and the CP3 group for its warm hospitality during the completion
of this work. We are also indebted to Barbara Mele 
for invaluable suggestions.

\end{document}